\documentclass{article}

\usepackage{microtype}
\usepackage{subfigure}
\usepackage{booktabs} 
\usepackage{graphicx}
\usepackage{hyperref}

\usepackage[accepted]{tom2023}

\usepackage{amsmath}
\usepackage{amssymb}
\usepackage{mathtools}
\usepackage{amsthm}

\usepackage[capitalize,noabbrev]{cleveref}

\theoremstyle{plain}

\theoremstyle{definition}

\theoremstyle{remark}

\makeatletter

\renewcommand{\fnum@figure}{\thefigure}
\renewcommand{\thefigure}{\textbf{Figure} \arabic{figure}}
\renewcommand{\fnum@table}{\thetable}
\renewcommand{\thetable}{\textbf{Table} \arabic{table}}

\newcommand*{\fullref}[1]{ \hyperref[{#1}]{\text{#1}}}

\usepackage[textsize=tiny]{todonotes}
\usepackage[superscript,biblabel]{cite}

\icmltitlerunning{}

\begin{document}

\twocolumn[
\icmltitle{Unveiling structure-property correlations in ferroelectric Hf\(_{0.5}\)Zr\(_{0.5}\)O\(_2\) films using variational autoencoders}

\icmlsetsymbol{equal}{*}

\begin{icmlauthorlist}
\icmlauthor{Kévin Alhada-Lahbabi}{yyy}
\icmlauthor{Brice Gautier}{yyy}
\icmlauthor{Damien Deleruyelle}{yyy}
\icmlauthor{Grégoire Magagnin}{yyy}\\

$^1$INSA Lyon, CNRS, Ecole Centrale de Lyon, Université Claude Bernard Lyon 1, CPE Lyon, INL, UMR5270, 69621 Villeurbanne, France.
\end{icmlauthorlist}

\icmlaffiliation{yyy}{}

\icmlcorrespondingauthor{kevin.alhada-lahbabi@insa-lyon.fr, gregoire.magagnin@ec-lyon.fr}

\icmlkeywords{Ferroelectrics, Phase-field, Machine learning, VAE, Hafnium Oxide}

\vskip 0.3in


\printAffiliationsAndNotice{} 

\begin{abstract}
While Hf\(_{0.5}\)Zr\(_{0.5}\)O\(_2\) (HZO) thin films hold significant promise for modern nanoelectronic devices, a comprehensive understanding of the interplay between their polycrystalline structure and electrical properties remains elusive.
Here, we present a novel framework combining phase-field (PF) modeling with Variational Autoencoders (VAEs) to uncover structure-property correlations in polycrystalline HZO. Leveraging PF simulations, we constructed a high-fidelity dataset of \(P\)-\(V\) loops by systematically varying critical material parameters, including grain size, polar grain fraction, and crystalline orientation. The VAEs effectively encoded hysteresis loops into a low-dimensional latent space, capturing electrical properties while disentangling complex material parameters interdependencies.
We further demonstrate a VAE-based inverse design approach to optimize \(P\)-\(V\) loop features, enabling the tailored design of device-specific key performance indicators (KPIs), including coercive field, remanent polarization, and loop area. The proposed approach offers a pathway to systematically explore and optimize the material design space for ferroelectric nanoelectronics. 
\end{abstract}
]

\section{Introduction}

Ferroelectric materials have been foundational in information and communication technologies for decades, with the first commercial FeRAMs (Ferroelectric Random-Access Memories) emerging in the 1990s. These perovskite-based devices demonstrated low-voltage, energy-efficient switching, consuming only 1-10 fJ/bit—up to three orders of magnitude lower than other memory technologies \cite{Khan2020}. However, their scalability has been limited to the 130 nm technology node, and their chemical incompatibility with CMOS technology hindered further advancements \cite{Takashima2011}. In contrast, fluorite-structured ferroelectrics, such as Hf\(_{0.5}\)Zr\(_{0.5}\)O\(_2\) (HZO), have emerged as promising candidates for next-generation devices due to their compatibility with CMOS \cite{Mulaosmanovic2021} and their demonstrated ferroelectricity in ultra-thin films (below a thickness of 2 nm) \cite{Cheema2020}. 

Despite these advantages, HZO-based ferroelectric devices face critical challenges, including wake-up, fatigue, and imprint. Wake-up refers to the gradual improvement in polarization states during initial cycling, while fatigue describes the eventual degradation of these states, reducing the device’s endurance and reliability \cite{Fengler2019}. Imprint, characterized by the shift in polarization hysteresis loops due to charge trapping or internal electric fields, further exacerbates variability and impacts long-term stability. These phenomena are intricately linked to nanoscale structural features such as grain boundaries, phase heterogeneities, and local defects, resulting in inconsistent performance across devices even under identical processing conditions.

To address these challenges, a multi-scale characterization framework is essential, bridging nanoscale material behavior with macroscopic device performance. Such an approach can correlate structural features like grain size, interfaces, and phase composition with device metrics such as endurance, retention, and coercive field. The complexity and non-linearity of these relationships pose significant challenges for conventional analytical methods. Recent advances in unsupervised machine learning, have demonstrated the potential to uncover latent physical mechanisms through variational autoencoders, offering transformative solutions for linking material behavior across scales \cite{Kalinin2023, Liu2024}.

VAEs are particularly well-suited for handling high-dimensional, multi-scale data. They compress complex datasets into a low-dimensional latent space, uncovering hidden patterns and key variables driving material behavior \cite{Kalinin2022, Liu2022}. This allows for the identification of critical correlations between nanoscale properties and macroscopic performance, such as the impact of grain size distribution on fatigue resistance or the role of phase heterogeneity in wake-up behavior. Furthermore, VAEs enable predictive modeling by exploring the latent space to identify optimal configurations for desired device properties \cite{Doersch2016}, such as finding initial film parameters that maximize remnant polarization for increasing a device memory window and/or minimize the coercive fields to reduce the device energy consumption \cite{Liu2021, Raghavan2024}.

The optimal film parameters for ferroelectric memory devices depend heavily on the specific device architecture and target application. For example, in Ferroelectric Field-Effect Transistors (FeFETs), minimizing leakage currents and optimizing the interface quality between the ferroelectric layer and the channel is critical for achieving high endurance and retention \cite{reviewfefet}. Similarly, for neuromorphic applications, controlling the polarization dynamics to attain stable, multi-level resistance states is crucial for mimicking synaptic behavior \cite{synapse,neuro1}. By leveraging the latent space of VAEs, it becomes possible to tailor film properties to meet the specific performance requirements of each device type, enabling more targeted and efficient material optimization.

High-throughput phase-field (PF) simulations have become fundamental for understanding ferroelectric materials, particularly Hf\(_{0.5}\)Zr\(_{0.5}\)O\(_2\) (HZO)-based devices \cite{phasefieldchen, Chenbase, reviewPF}. Recent studies have modeled ferroelectric hysteresis in HZO \cite{FerroX_nopoly, Saha_landau_calibrating, Saha2020_hzo_nature_report, hzo_pf_90domains, 2DPfhZO, hZOpf_device, GNNHZO} and analyzed the effects of polycrystalline grain structures on electrical properties \cite{Genetic_HZO, 3D_hzo_grain}. Phase-field modeling has also been applied to investigate electric-field-induced phase transitions \cite{hzocrystallizationphasefield}, spacer layer impacts on domain dynamics \cite{chenhzo}, and the role of oxygen vacancies in electrical behavior \cite{phasefieldOxygen, phasefieldOxygen1, phasefieldOxygen2}. 
PF modeling thus enables the generation of high-fidelity datasets capturing polarization evolution and field-induced dynamics in HZO thin films. 

In this work, we first leverage PF modeling to simulate a wide range of material parameter distributions, such as variations in grain size, phase fraction, and defect distributions, to generate rich datasets for training the VAE. Next, we demonstrate how the VAE latent space disentangles the complex interplay of these material parameters and their impact on HZO's electrical properties. Finally, we illustrate the inverse design of ferroelectric hysteresis properties using the VAE framework. This integration of PF simulation and VAE facilitates systematic exploration of the material design space, offering valuable guidance for experimental efforts to optimize functional properties.

\section{Results}

\subsection{Material Parameter space}

In this study, our primary objective is to investigate how the interaction between key material parameters governing polycrystalline structures influences the electrical properties of HZO, utilizing PF modeling and VAE. Specifically, we aim to explore the impact of four critical parameters that dictate the crystalline arrangement: the average grain size (\(D_G\)), the proportion of orthombic polar to monoclinic non-polar grains (\(\nu_{ortho}\)), the preferential orientation of the crystallographic axis (\(\mu_\theta\)), and the dispersion of grain orientations relative to this axis (\(\sigma_\theta\)).

\begin{figure*}[!h]
    \centering
    \includegraphics[width=.9\linewidth]{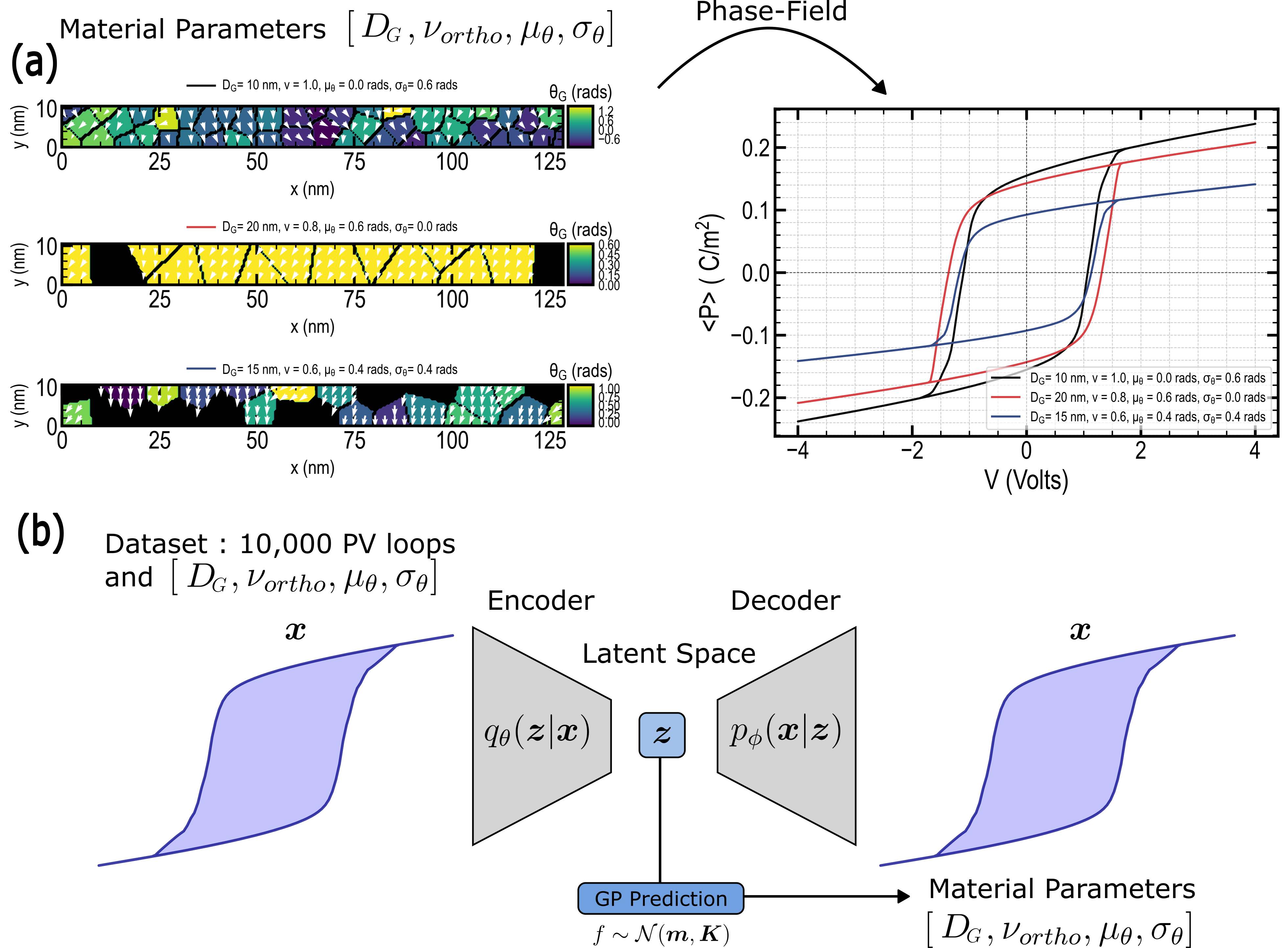}
    \caption{\textbf{Schematic Workflow for Using VAE to Disentangle Material Parameters' Influence on HZO Ferroelectric Hysteresis.}  
\textbf{(a)} Three polycrystalline structures, each with a unique set of material parameters—grain size (\(D_G\)), grain orientation dispersion (\(\sigma_\theta\)), preferential crystalline orientation (\(\mu_\theta\)), and polar grain fraction (\(\nu_{\text{ortho}}\))—and their corresponding \(P\)-\(V\) loops from PF modeling. Grain orientation is color-coded, with white arrows representing polarization direction, and black regions corresponding to non-polar grains and grain boundaries. \textbf{(b)} The VAE workflow, encoding \(P\)-\(V\) loops into a 2D latent space and decoding them. GPs predict the material parameters from the latent variables, connecting the latent space to the original material property space.}
    \label{fig:figure1}
\end{figure*}

In \ref{fig:figure1}a, three examples of \(10 \, \text{nm} \times 128 \, \text{nm}\) 2D polycrystalline structures, generated via Voronoi tessellation, are presented, each corresponding to a distinct combination of material parameters \([D_G, \nu_{\text{ortho}}, \mu_\theta, \sigma_\theta]\). These structures exhibit various grain arrangements based on the average grain size \(D_G\) (here ranging from 10 nm to 20 nm), encompassing both equiaxed and columnar grains. The polarization vector field, initialized in a downward orientation, is represented by white arrows, illustrating the interplay between the fiber texture \(\mu_\theta\) and the grain orientation dispersion \(\sigma_\theta\), as well as their effect on the grain orientation distribution \(\theta_G\). Furthermore, non-polar grains are depicted in black, indicating a polarization fraction ranging from 100\% to 60\%. 
The corresponding ferroelectric hysteresis loops, obtained from PF simulations, are shown alongside, highlighting the influence of crystalline parameters on the electrical properties of HZO. Specifically, the coercive field \(E_c\), remanent polarization \(P_r\), loop area, and saturation polarization \(P_S\) exhibit significant variation with changes in the polycrystalline structure, revealing a complex relationship between the material parameters and the resulting \(P\)-\(V\) loops.

To elucidate the structure-property correlations in ferroelectric Hf\(_{0.5}\)Zr\(_{0.5}\)O\(_2\), we employ a VAE, as illustrated in \ref{fig:figure1}b. A dataset of 10,000 PF \(P\)-\(V\) loops is generated by sampling various combinations of material parameters \([D_G, \nu_{\text{ortho}}, \mu_\theta, \sigma_\theta]\) (see Methods). The VAE takes the polarization points of the ferroelectric hysteresis loops as input \(\boldsymbol{x}\), encodes them into a 2D latent space, and decodes them to reconstruct the original \(P\)-\(V\) loops. Subsequently, a predictive model, chosen here as Gaussian Processes (GPs), is trained in a supervised manner to infer the material parameters from the latent variables, thereby linking the latent space to the original material parameter space.

\subsection{VAE Latent Space Analysis}

\begin{figure*}[!h]
    \centering
    \includegraphics[width=\linewidth]{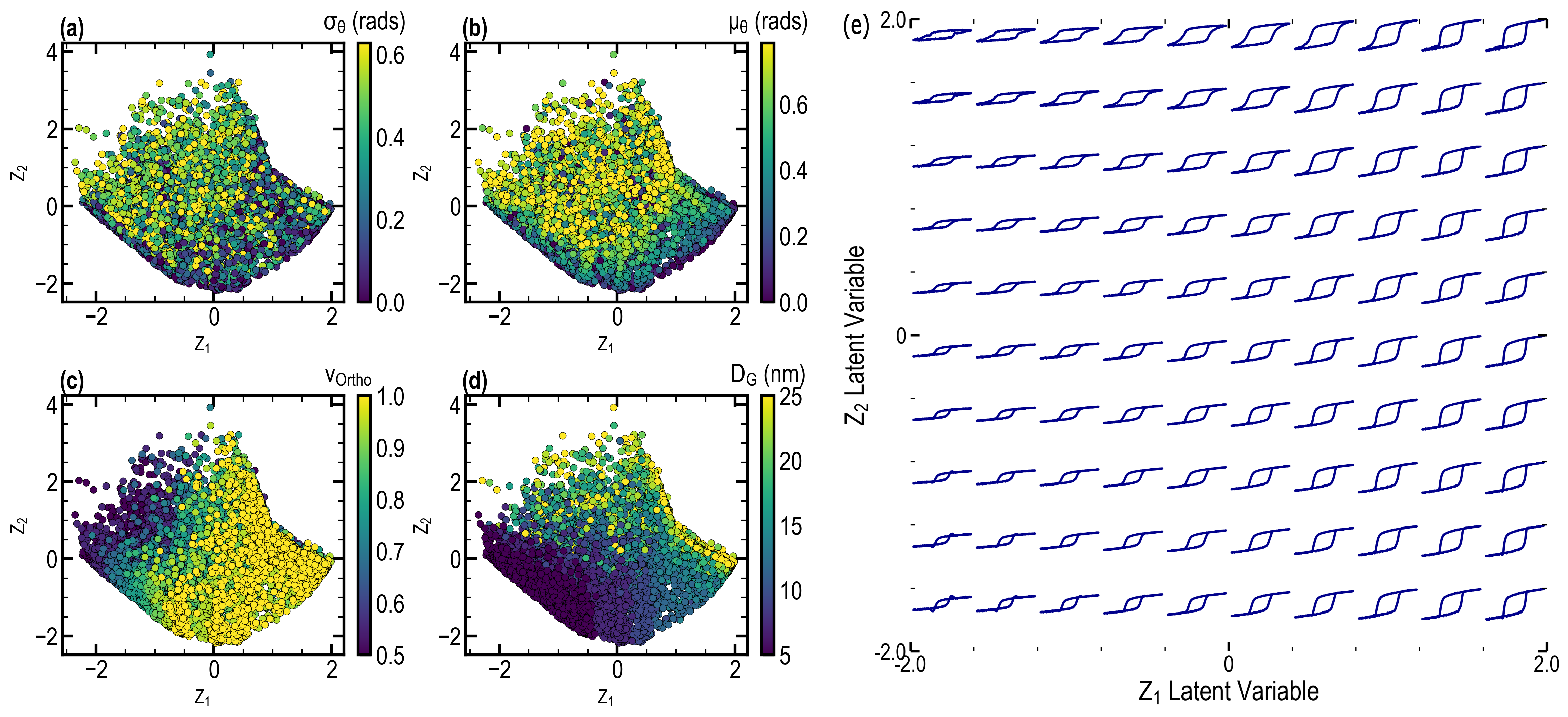}
    \caption{\textbf{Representation of the Ferroelectric Hysteresis Dataset in the VAE Latent Space.} 
The 2D latent space (\(Z_1, Z_2\)) of the VAE is used to represent the latent encoding of ferroelectric hysteresis data. Each point in the latent space corresponds to a hysteresis loop, with the points colored by the material properties used to generate the respective polarization-voltage (P-V) loops. The material properties include: \textbf{(a)} grain orientation dispersion (\(\sigma_\theta\)), \textbf{(b)} preferential crystalline orientation (\(\mu_\theta\)), \textbf{(c)} polar grain fraction (\(\nu_{\text{ortho}}\)), and \textbf{(d)} average grain size (\(D_G\)). Additionally, \textbf{(e)} the latent manifold is constructed by decoding points sampled uniformly within the \([-2, 2] \times [-2, 2]\) square in the latent space.}
    \label{fig:figure2}
\end{figure*}

After training is complete, the 10.000 \(P\)-\(V\)  loops comprising the datasets are represented in the VAE latent space in \ref{fig:figure2}, based on the two latent variables \(Z_1,Z_2\).
Here, the latent representation of the ferroelectric hysteresis is color-coded based on their grain orientation dispersion (\(\sigma_\theta\)) (\ref{fig:figure2}a),  preferential crystalline orientation (\(\mu_\theta\)) (\ref{fig:figure2}b),  polar grain fraction (\(\nu_{\text{ortho}}\)) (\ref{fig:figure2}c), and \ average grain size (\(D_G\)) (\ref{fig:figure2}d), allowing to elucidate the correlation between the material properties and VAE encoding. 
Additionally, the latent manifold resulting from the reconstruction of a continuous square of dimensions \([-2, 2] \times [-2, 2]\) in the latent space is shown in \ref{fig:figure2}e offering a continuous representation of the encoded hysteresis behaviors. This manifold highlights the VAE's structured organization of the latent space, where \(P\)-\(V\) loops are predominantly arranged based on their remanent polarization, saturation polarization, coercive field, loop area, and shape across the dataset.

\begin{figure}[!h]
    \centering
    \includegraphics[width=\linewidth]{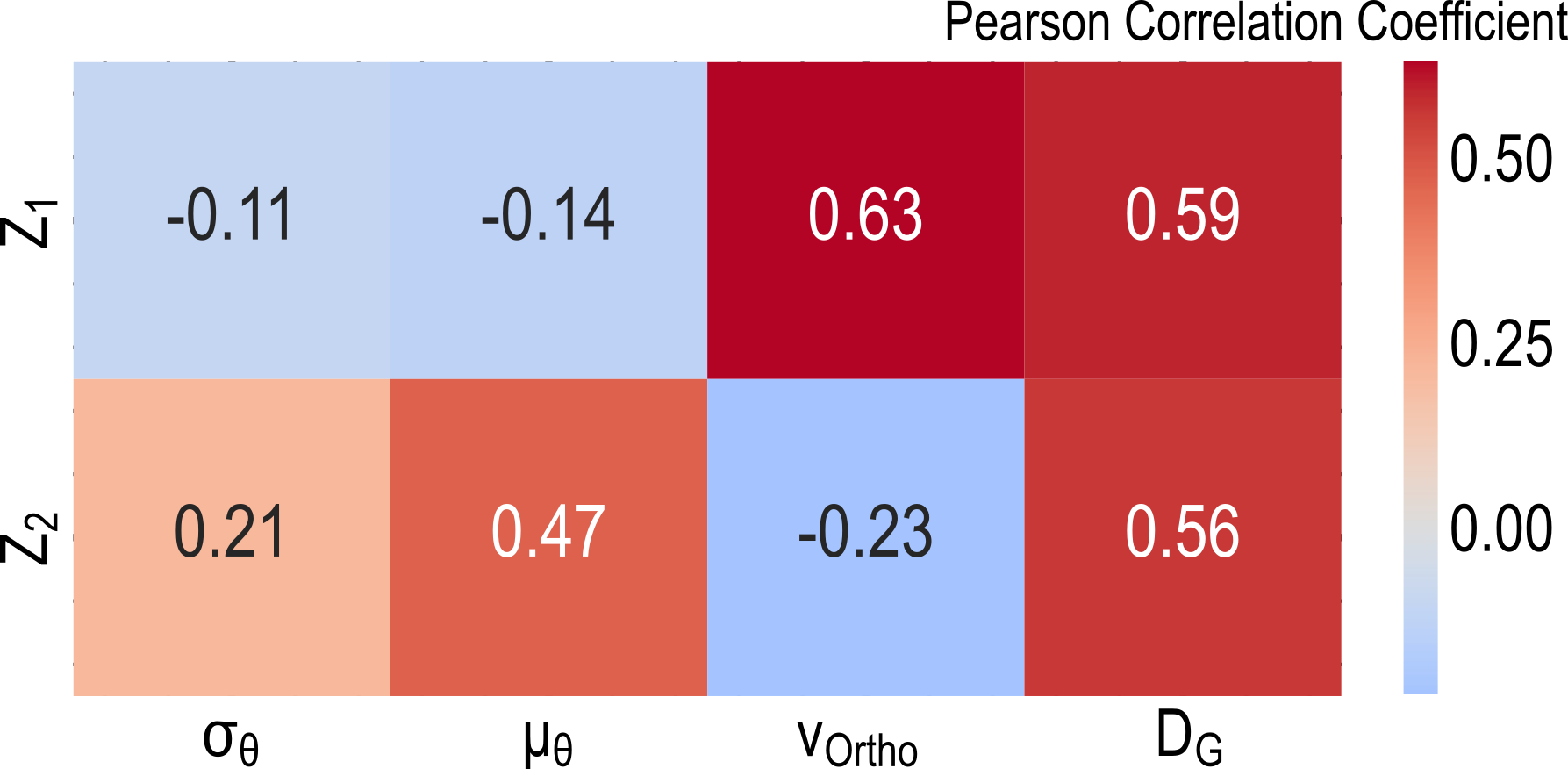}
    \caption{\textbf{Pearson Correlation Coefficient Between Latent Variables and Material Properties.}  
Pearson correlation coefficients between the 2D latent variables (\(Z_1, Z_2\)) derived from the VAE encoding and the material properties associated with the ferroelectric hysteresis loops: grain orientation dispersion (\(\sigma_\theta\)), preferential crystalline orientation (\(\mu_\theta\)), polar grain fraction (\(\nu_{\text{ortho}}\)), and average grain size (\(D_G\)). }
    \label{fig:figure3}
\end{figure}

The VAE representation effectively captures the relationship between material parameters and \(P\)-\(V\) loop features, as organized within the latent space. For instance, the grain orientation dispersion (\(\sigma_\theta\)) shows minimal correlation with the latent variables, with no discernible clusters or organization in the latent space. This suggests a limited influence on the \(P\)-\(V\) loop characteristics compared to other material parametesr, further supported by its relatively low Pearson correlation coefficients (\(r_{Z_1}=-0.11,r_{Z_2}=0.21\)), as depicted in \ref{fig:figure3}(see Methods). 
In contrast, the preferential crystallographic orientation (\(\mu_\theta\)) exhibits a stronger correlation, particularly with the \(Z_2\) latent variable (\(r_{Z_2}\)=0.47). This dependency is evident in the latent manifold, where \(\mu_\theta\) values near 0 (corresponding to a vertically aligned c-axis) are concentrated in the lower quadrant, resulting in more open \(P\)-\(V\) loops with higher coercive fields. This behavior arises from the enhanced coupling between polarization and the vertical electric field when the c-axis aligns with the out-of-plane direction.
The polar fraction (\(\nu_{\text{ortho}}\)) demonstrates a strong correlation with \(Z_1\) (\(r_{Z_1}\)=0.63), as reflected in the latent manifold, where increases in \(\nu_{\text{ortho}}\) correspond to higher remanent and saturation polarization along the \(Z_1\) axis. 
Finally, the average grain size (\(D_G\)) shows significant correlation with both \(Z_1\) and \(Z_2\) (\(r_{Z_1}\)= 0.59 and \(r_{Z_2}\)= 0.56, respectively). Larger grain sizes in the latent space are associated with higher remanent and saturation polarization, as well as increased loop opening and coercive field values.

\begin{figure*}[!h]
    \centering
    \includegraphics[width=\linewidth]{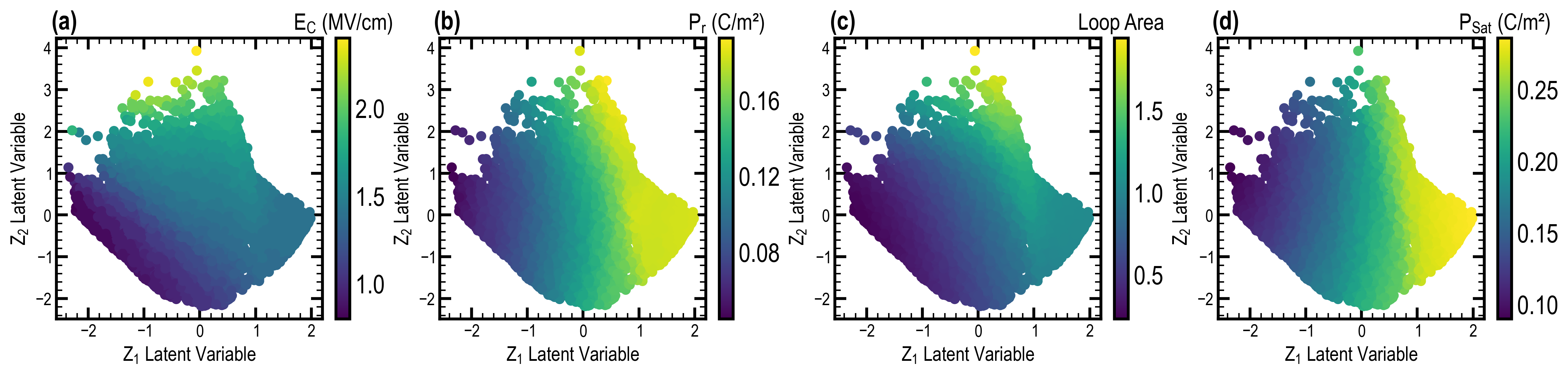}
    \caption{\textbf{Representation of the evolution  Ferroelectric Hysteresis Properties within the VAE Latent.}  
The 2D latent space (\(Z_1, Z_2\)) derived from the VAE encoding is visualized with color-coded subplots corresponding to key ferroelectric hysteresis loop properties: \textbf{(a)} coercive field (\(E_c\)), \textbf{(b)} remanent polarization (\(P_r\)), \textbf{(c)} hysteresis loop area, and \textbf{(d)} saturation polarization (\(P_s\)). }
    \label{fig:figure4}
\end{figure*}

To further explore the VAE latent representation, the latent space is visualized in \ref{fig:figure4}, with color coding based on features extracted from the reconstructed \(P\)-\(V\) loops. Specifically, the coercive field (\(E_c\)) (\ref{fig:figure4}a), remanent polarization (\(P_r\)) (\ref{fig:figure4}b), hysteresis loop area (\ref{fig:figure4}c), and saturation polarization (\(P_s\)) (\ref{fig:figure4}d) were derived from the decoded latent points. 
Remarkably, these \(P\)-\(V\) features exhibit a smooth and continuous variation across the latent space, highlighting the VAE's ability to disentangle and organize key hysteresis properties effectively. This continuity not only underscores the model's efficiency in representing \(P\)-\(V\) loop features but also opens avenues for generating and inversely designing \(P\)-\(V\) loops with specified properties, leveraging the latent space's structured representation.

\begin{figure*}[!h]
    \centering
    \includegraphics[width=\linewidth]{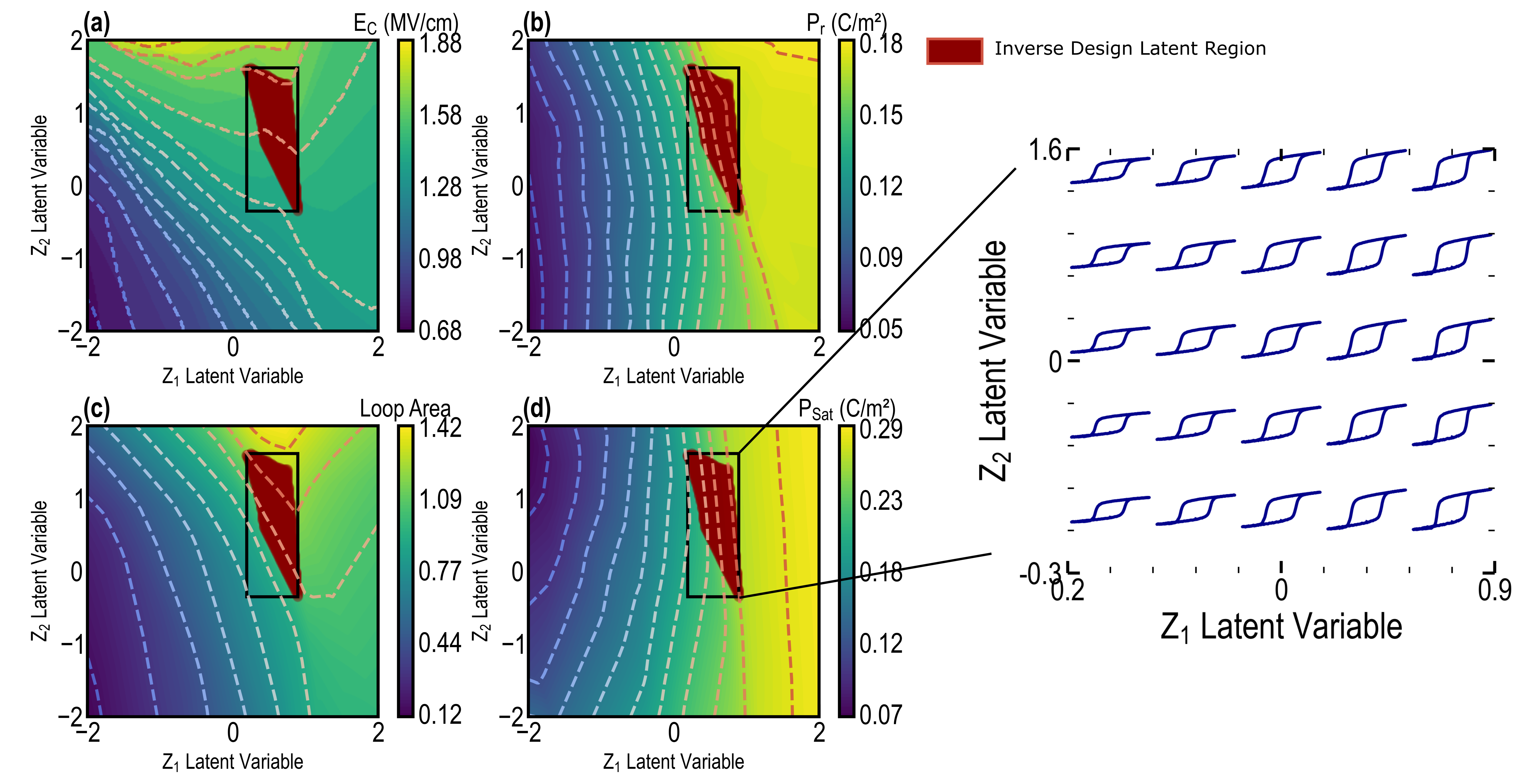}
   \caption{\textbf{Inverse Design of \(P\)-\(V\) Features from VAE Latent Space Decoding.}   
Ferroelectric hysteresis properties, including \textbf{(a)} coercive field (\(E_c\)), \textbf{(b)} remanent polarization (\(P_r\)), \textbf{(c)} loop area, and \textbf{(d)} saturation polarization (\(P_s\)), obtained by decoding the \([-2, 2] \times [-2, 2]\) latent space. The red region highlights latent points whose decoded \(P\)-\(V\) loops satisfy the inverse design criteria for \(E_c\), loop area, \(P_r\), and \(P_s\). The manifold representation within the \([0.2, 0.9] \times [-0.3, 1.6]\) black rectangle illustrates ferroelectric hysteresis loops that meet the specified design requirements.}
    \label{fig:figure5}
\end{figure*}

\subsection{VAE for the Inverse Design of Ferroelectric Hysteresis}

Using the proposed VAE-based approach, the inverse design of ferroelectric hysteresis with targeted properties becomes possible through the generation of realistic \(P\)-\(V\)  loops from the learned latent space. Specifically, \(P\)-\(V\)  loops with a specified range for the coercive field, remanent polarization, saturation polarization, and loop area, can be obtained, fulfilling the requirement specific to different applications.

To demonstrate the potential of the inverse design approach, we seek to identify a ferroelectric hysteresis loop in the latent space with specific feature characteristics. The desired properties are arbitrarily chosen as follows: a coercive field (\(E_c\)) between 1.4 and 1.6 MV/cm, remanent polarization (\(P_r\)) between 0.15 and 0.25 C/m², saturation polarization (\(P_s\)) between 0.20 and 0.26 C/m², and a hysteresis loop area ranging from 1 to 1.5 a.u.

To locate a region in the VAE latent space that meets these criteria, we continuously decode the \([-2, 2] \times [-2, 2]\) square region and extract the corresponding \(P\)-\(V\)  loop characteristics. This results in a continuous mapping from the latent space to the \(P\)-\(V\)  properties, as illustrated in \ref{fig:figure5}. By analyzing this mapping, we can identify the latent region that satisfies the specified requirements. In this case, the region of interest is highlighted in red on the 2D latent space.
Furthermore, a representation of the generated \(P\)-\(V\)  loop manifold in the vicinity of this region, within the latent coordinates \([0.2, 0.9] \times [-0.3, 1.6]\), is shown in \ref{fig:figure5}e, where the recovered ferroelectric hysteresis loops fulfill the design specifications.

Finally, the material parameters corresponding to the ferroelectric hysteresis generated from the VAE latent space must be recovered to complete the inverse design process. While the VAE can effectively generate samples from latent regions different from the training points to a certain extent, reconstructing the material parameters requires the mapping of the latent space to the material parameters \([D_G, \nu_{\text{ortho}}, \mu_\theta, \sigma_\theta]\). This step may be challenging, especially if the material parameter do not exhibit a significant correlation with the latent variables.

\begin{figure*}[!h]
    \centering
    \includegraphics[width=\linewidth]{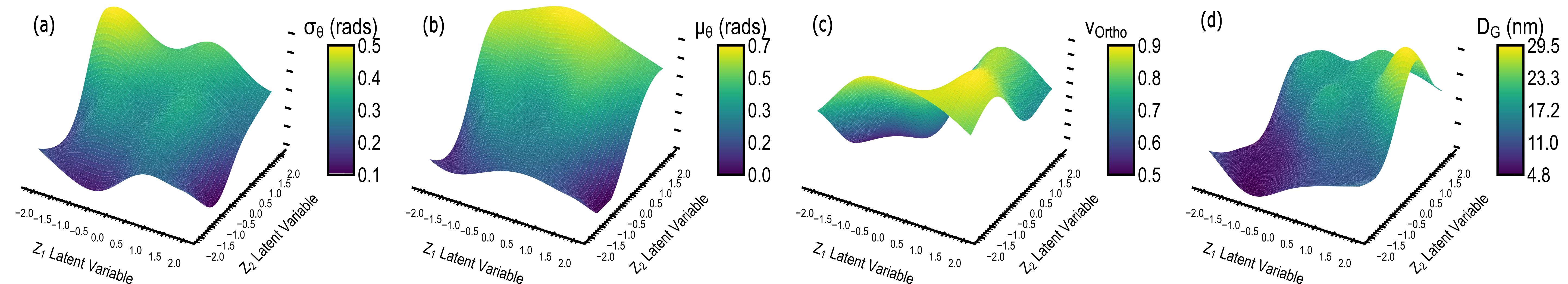}
    \caption{\textbf{GP Predictions of Material Properties in the VAE Latent Space.}  
GP regression of material properties from the 2D latent space (\(Z_1, Z_2\)) derived from the VAE. Predictions were made across a \([-2, 2] \times [-2, 2]\) square in the latent space for the \textbf{(a)} grain orientation dispersion (\(\sigma_\theta\)), \textbf{(b)} preferential crystalline orientation (\(\mu_\theta\)), \textbf{(c)} polar grain fraction (\(\nu_{\text{ortho}}\)), and \textbf{(d)} average grain size (\(D_G\)).}
    \label{fig:figure6}
\end{figure*}

Here, we employ GPs to map the latent variables \(Z_1, Z_2\) to the material parameters \([D_G, \nu_{\text{ortho}}, \mu_\theta, \sigma_\theta]\) (see Methods):
\[
 f_{\text{GP}}\left( [ Z_1, Z_2] \right) = [D_G, \nu_{\text{ortho}}, \mu_\theta, \sigma_\theta]
\]
where \(f_{\text{GP}}\) represents the GP model that predicts the material parameters from the latent variables. Notably, a separate GP model was trained for each material parameter.
The GP prediction of the material parameters over the \([-2, 2] \times [-2, 2]\) latent space is depicted in \ref{fig:figure6}.

The GPs can be further utilized to predict the material parameters corresponding to the latent region that satisfies the design requirements. For instance, by inputting the latent point \(\boldsymbol{Z} = (0.8, 0.5)\) which is present in this inverse design region, the GP model predicts the following material parameters: average grain size \(D_G = 16.2 \, \text{nm}\), polar fraction \(\nu_{\text{ortho}} = 0.87\), preferential crystalline orientation \(\mu_\theta = 0.54 \, \text{rads}\), and grain orientation dispersion \(\sigma_\theta = 0.38 \, \text{rads}\).
It is important to note that the predictions made by the GPs are inherently associated with uncertainty, allowing for the determination of confidence intervals for the predicted material parameters. These intervals provide bounds on the reliability of the predictions. The standard deviations associated with the predictions are as follows: \(\sigma_{D_G} = 3.4 \, \text{nm}\), \(\sigma_{\nu_{\text{ortho}}} = 0.09\), \(\sigma_{\mu_\theta} = 0.15 \, \text{rads}\), and \(\sigma_{\sigma_\theta} = 0.15 \, \text{rads}\).

\begin{figure}[!h]
    \centering
    \includegraphics[width=\linewidth]{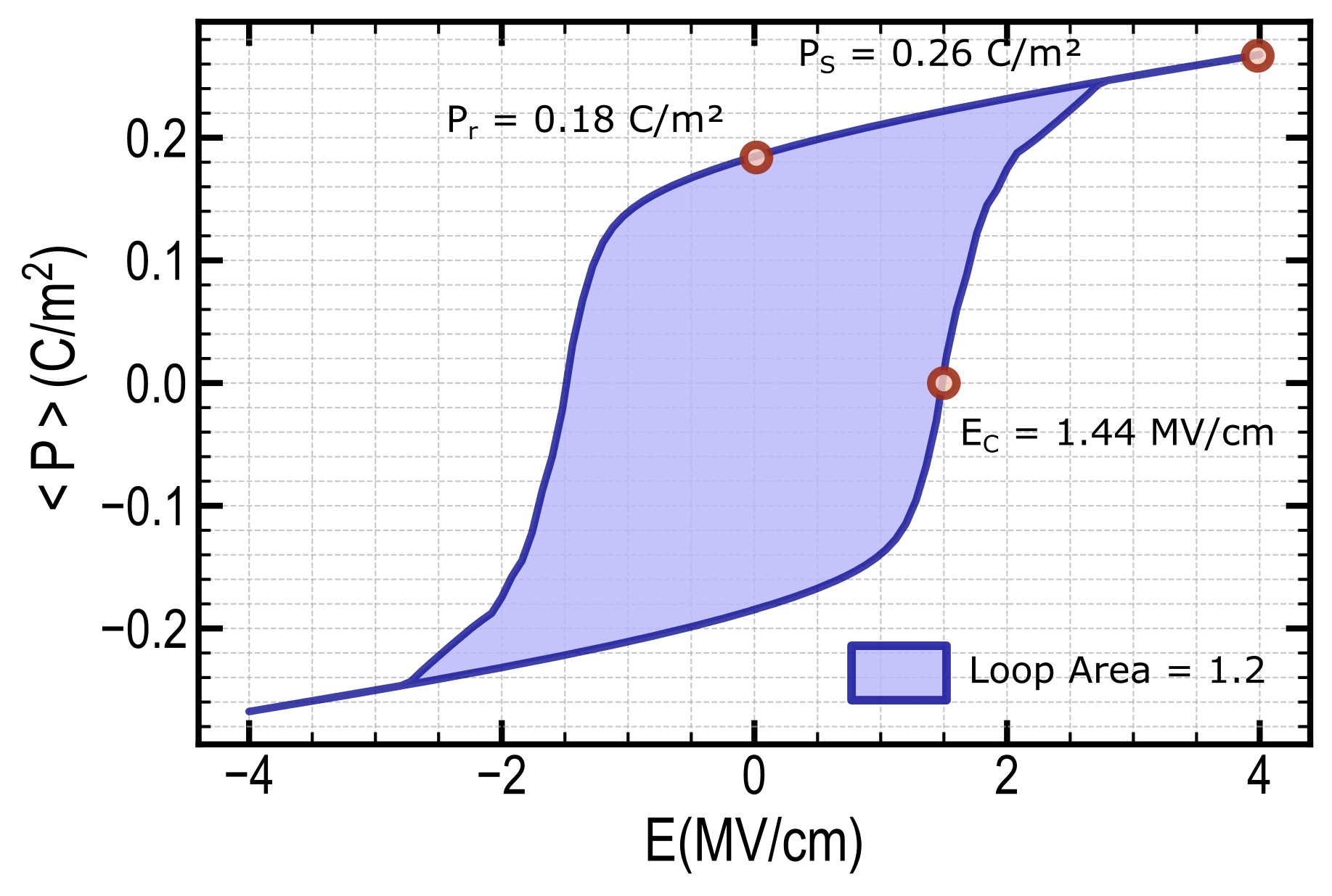}
    \caption{\textbf{\(P\)-\(V\) Loop Modeled from Voronoi Structure Based on Material Parameters from VAE Inverse Design.}  
The ferroelectric hysteresis \(P\)-\(V\) loop is simulated using a PF model, with material parameters derived from the VAE inverse design process. The generated loop meets the predefined design requirements, reflecting the influence of the material parameters on the \(P\)-\(V\) characteristics.}
    \label{fig:figure7}
\end{figure}

Finally, the recovered material parameters are utilized to generate the polycrystalline structure and the corresponding phase-field \(P\)-\(V\) loop, as presented in \ref{fig:figure7}. The simulated \(P\)-\(V\) loop demonstrates the following characteristics: coercive field \(E_c = 1.44 \, \text{MV/cm}\), loop area of 1.2 a.u., remanent polarization \(P_r = 0.18 \, \text{C/m}^2\), and saturation polarization \(P_s = 0.26 \, \text{C/m}^2\), all of which align with the specified design criteria. These outcomes underscore the successful implementation of the inverse design methodology in achieving the targeted \(P\)-\(V\) loop properties.

\section{Discussion}

In this study, we introduced an approach based on VAEs and PF modeling of polycrystalline Hf\(_{0.5}\)Zr\(_{0.5}\)O\(_2\) to elucidate the interplay between the polycrystalline structure and the electrical properties of the material.

The usefulness of the proposed VAE-based inverse design approach lies in its ability to identify key material parameters that directly influence the performance of ferroelectric devices. By mapping the latent space to specific features of the P-V loop, such as coercive field, remnant polarization, and loop area, the framework enables the optimization of these properties to meet specific device requirements. For example, in FeFETs, minimizing the coercive field and maximizing remnant polarization is critical to reducing energy consumption and increasing memory window \cite{reviewfefet}. This inverse design framework allows for targeted exploration of the polycrystalline HZO parameter space, optimizing variables such as grain size, polar grain fraction, and crystallographic orientation to achieve desired electrical properties. The ability to recover and predict material parameters from the latent space provides a systematic pathway to tailor the polycrystalline structure for device-specific KPIs. For instance, reducing imprint voltage or mitigating fatigue effects can be targeted by precisely adjusting the grain size distribution or polar fraction in the functional film.

VAEs are a powerful tool for disentangling the complex relationships between material parameters in the PF modeling of polycrystalline HZO. This method provides valuable insights into the interdependencies of these parameters. Furthermore, we introduced an inverse design framework that leverages the continuous generative capabilities of the VAE, combined with GPs, to design ferroelectric hysteresis loops with specific properties and recover the corresponding material parameters governing the polycrystalline structure.

The Landau coefficients were arbitrarily chosen for this study and remain constant across the dataset. It is crucial to note that the KPIs, such as remanent and saturation polarization magnitudes, are significantly influenced by the Landau coefficients. While this study primarily focuses on the impact of the polycrystalline structure, it is also possible to investigate the interplay between the polycrystalline structure and the Landau coefficients with the VAE approach. Additionally, the inverse design approach could be extended to calibrate these Landau coefficients, which remains an actual challenge for polycrystalline HZO ferroelectrics.

However, the computational demands of generating large PF simulation datasets for VAE training remain a challenge. Although, integrating ML surrogates, such as graph neural networks or transfer learning models, can significantly accelerate data generation \cite{Alhada_GNN, AlhadaNPJ, MontesdeOcaZapiain2021,alhadatransfer}. These ML surrogates could be incorporated into the VAE framework to efficiently generate a more diverse and comprehensive dataset, potentially extending to 3D PF simulations and enabling the inclusion of additional parameters, such as film thickness, or dielectric layer properties, further enhancing the design space and predictive power of the framework.

Moreover, the flexibility of this framework extends beyond hysteresis loop properties. It can be adapted to optimize other material response characteristics, such as butterfly loop behavior, fatigue resistance, or frequency-dependent switching dynamics. This broad applicability makes the approach highly valuable for advancing ferroelectric devices across diverse applications, from memory and logic to energy storage and neuromorphic computing.

In conclusion, we introduced an ML framework combining VAEs and phase-field modeling to provide a robust and versatile tool for elucidating structure-property relationships in ferroelectric HZO thin films. By enabling the inverse design of material parameters tailored to device-specific KPIs, this approach paves the way for systematic and efficient optimization of next-generation ferroelectric devices.

\section{Methods}

\subsection{Phase-Field Modeling}

In this study, we utilized PF simulations to explore how material parameters of polycrystalline Hf\(_{0.5}\)Zr\(_{0.5}\)O\(_2\) influence its electrical properties, building on the framework presented in recent research efforts \cite{Alhada_GNN,Kalhada_AEM}. Key factors analyzed were the average grain diameter (\(D_{G}\)), the fraction of polar orthorhombic grains (\(\nu_{ortho}\)), the preferential crystalline axis orientation (\(\mu_{\theta}\)), and the dispersion of grain orientations (\(\sigma_{\theta}\)).

Simulations were conducted on a 2D grid (\(10 \, \text{nm} \times 128 \, \text{nm}\)) with a uniform spacing of \(\Delta x = \Delta y = 0.5 \, \text{nm}\). Polycrystalline structures were generated via Voronoi tessellation based on specified average grain diameters \(D_{G}\). Grain boundaries, sampled uniformly between \(1 \, \text{nm}\) and \(2 \, \text{nm}\), align with values commonly reported in the literature~\cite{ebsd_grain_hzo,Exp_Lit_grain_4_5_15,Exp_Lit_grain_4_10_20}.

Ferroelectric grain orientations were modified by assigning a random angle \(\theta_G\) to each grain, which defines the orientation of its polarization axis relative to the vertical direction. The corresponding rotation matrix \(\hat{R}\) was used to compute the free energy in the local crystalline coordinate system. This matrix transforms a global vector \(\boldsymbol{r} = (x, y)\) into its local representation \(\boldsymbol{r}^L = (x^L, y^L)\) as \(\boldsymbol{r}^L = \hat{R} \boldsymbol{r}\):

\[
\hat{R} = 
\begin{bmatrix}
\cos(\theta_G(\boldsymbol{r})) & -\sin(\theta_G(\boldsymbol{r})) \\
\sin(\theta_G(\boldsymbol{r})) & \cos(\theta_G(\boldsymbol{r}))
\end{bmatrix}.
\]

To analyze the interplay between general crystalline texturing \(\mu_\theta\) (e.g., preferential orientations such as [001] or [111]) and the dispersion of grain orientations \(\sigma_\theta\), grain angles were sampled from a Gaussian distribution with a mean orientation \(\mu_\theta\) and standard deviation \(\sigma_\theta\) (\(\theta_G(\boldsymbol{r})) \sim \mathcal{N}(\mu_\theta, \sigma_\theta)\)).

Additionally, the influence of the distribution of polar and non-polar grains within the crystalline structure was considered by varying the ratio of orthorhombic ferroelectric grains \(\nu_{ortho}\). For each polycrystalline structure, a fraction of \(1-\nu_{ortho}\) grains was randomly selected to be non-polar, representing the non-ferroelectric proportion of the polycrystalline system.

The evolution of the spontaneous polarization $P_i(\boldsymbol{r},t)$ is given by the time-dependent Landau Ginzburg (TDGL) equation\cite{Chenbase} :
\begin{equation}
    \frac{\partial P_i(\boldsymbol{r},t)}{\partial t} = -L_0\frac{\delta F}{\delta P_i(\boldsymbol{r},t)}, \quad (i=1,2)
\end{equation}
where  $L_0$ is a kinetic coefficient and $F$ is the total free energy, which includes the different energetic contributions,
\begin{equation}
    F = \int_V (f_{\text{Landau}}+f_{\text{grad}}+f_{\text{electric}}+f_{\text{elastic}})dV
\end{equation}
where  \(f_{\text{Landau}}\), \(f_{\text{grad}}\), \(f_{\text{electric}}\), and \(f_{\text{elastic}}\) are the bulk, gradient, electric, and elastic free energy density, respectively.


In the polar Hf\(_{0.5}\)Zr\(_{0.5}\)O\(_2\) grains, the polarization was assumed to be directed along the c-axis of the crystal, as commonly considered in HZO PF literature\cite{Genetic_HZO,hzocphasefield}, with each grain's orientation \(\theta_G\) taken into account. In this configuration, the bulk free energy is described by the Landau expansion:
\[
f_{\text{Landau}} = \alpha_1 P_1^2 + \alpha_{11} P_1^4 + \alpha_{111} P_1^6
\]
where \(P_1\) represents the out-of-plane polarization aligned with the grain c-axis, and \(\alpha_1\), \(\alpha_{11}\), and \(\alpha_{111}\) are the Landau coefficients.

The energy contribution from domain walls is described by the gradient energy, assumed to be isotropic. It can be expressed as:
\begin{align}
    f_{\text{grad}} &= G_{11} \left( P_{1,1}^2 + P_{2,2}^2 \right) 
    + G_{12} \left( P_{1,1}P_{2,2} \right) \\ \nonumber
    &+ \frac{1}{2} G_{44} [ \left( P_{1,2} + P_{2,1} \right)^2 
     ] + \frac{1}{2} G'_{44} [ \left( P_{1,2} - P_{2,1} \right)^2 
     ].
\end{align}
where \(G_{ijkl}\) are the gradient energy coefficients, and \(P_{i,j}\) denotes the partial derivative \(\frac{\partial P_i}{\partial x_j}\).

The electrostatic energy is given by:
\[
f_{\text{electric}} = -P_i E_i - \frac{1}{2} \epsilon_0 \epsilon_r E_i E_j
\]
where \(E_i\) is the electric field, \(\epsilon_0\) is the vacuum permittivity, and \(\epsilon_r\) is the relative dielectric permittivity. The electric field \(E_i(\boldsymbol{r}) = -\frac{\partial V(\boldsymbol{r})}{\partial x_i}\) is derived from the electrostatic potential \(V\), which satisfies the electrostatic equilibrium equation:
\[
\frac{\partial}{\partial x_i} \left[ \epsilon_0 \epsilon_r(\mathbf{r}) \frac{\partial V(\mathbf{r})}{\partial x_j} \right] = \frac{\partial P_i(\mathbf{r})}{\partial x_i} - \rho(\mathbf{r})
\]
Here, \(\rho\) is the charge density, and \(\frac{\partial P_i}{\partial x_i}\) is the polarization charge density. The electrostatic equilibrium is solved using the fast Fourier transform iterative spectral method with periodic boundary conditions in the in-plane directions and Dirichlet boundary conditions in the out-of-plane direction \cite{chen_FFT_inho}.

The elastic energy density is given by:
\[
f_{\text{elastic}} = \frac{1}{2} C_{ijkl} \left( \epsilon_{ij}(\mathbf{r}) - \epsilon^0_{ij}(\mathbf{r}) \right) \left( \epsilon_{kl}(\mathbf{r}) - \epsilon^0_{kl}(\mathbf{r}) \right)
\]
where \(C_{ijkl}\) is the elastic stiffness tensor, \(\epsilon_{ij}(\mathbf{r})\) is the total strain, and \(\epsilon_{ij}^0(\mathbf{r}) = Q_{ijkl} P_k(\mathbf{r}) P_l(\mathbf{r})\) is the electrostrictive strain induced by the polarization, with \(Q_{ijkl}\) being the electrostrictive tensor.

The mechanical equilibrium equation is:
\[
C_{ijkl} \frac{\partial^2 u_k(\boldsymbol{r})}{\partial x_j \partial x_l} = C_{ijkl} \frac{\partial \epsilon_{kl}^0(\boldsymbol{r})}{\partial x_j}
\]
where \(u_k\) represents the mechanical displacements, determined using thin-film mechanical boundary conditions \cite{phasefieldchen, chenhzo}, with substrate and air layer thicknesses of \(12 \Delta x\) and \(2 \Delta x\), respectively.

To model the behavior of non-ferroelectric regions, such as non-polar grains and grain boundaries, the polarization was not governed by the TDGL equation. Instead, it was assumed to evolve as:
\[
P_i(\boldsymbol{r}, t) = \epsilon_0 (\epsilon_r - 1) E_i(\boldsymbol{r}, t), \quad (i=1,2)
\]
consistently with previous studies on polycrystalline ferroelectric PF models \cite{GB_fedeli}.

Ferroelectric hysteresis was obtained using a uniformly discretized voltage ramp with 400 steps, ranging from -4 to 4 volts. At each step, a constant voltage was applied for \(50\Delta t\). The electrostatic potential \(u_T\) at the top electrode was adjusted such that \(V(x_{\text{top}}, y, z) = u_T\), while the bottom electrode was grounded with \(V(0, y, z) = 0\) V. The average polarization \(<P(V)>\) was recorded at each step, resulting in 400 data points per P-V loop.

The Landau coefficients were selected as \(\alpha_1 = -7.79 \times 10^8\) C\(^{-2}\)m\(^2\)N, \(\alpha_{11} = 1.15 \times 10^{10}\) C\(^{-4}\)m\(^6\)N, and \(\alpha_{111} = -2.51 \times 10^{10}\) C\(^{-6}\)m\(^ {10}\)N. The dielectric permittivity for HZO was set as \(\epsilon_r^{G} = 30\) in the grains \cite{eps30}, and an arbitrary lower value of \(\epsilon_r^{GB} = 12\) was chosen for the grain boundaries. The remaining coefficients for the PF simulation were taken from Ref. \cite{chenhzo}. The time step was set as \(\Delta t = 0.06t_0\), where \(t_0 = 1/(\alpha_0L_0)\).

To investigate the interplay between the material parameters governing the polycrystalline structure, the average grain diameter \(D_G\), the polar orthorhombic fraction \(\nu_{\text{ortho}}\), crystalline fiber texture \(\mu_{\theta}\), and crystalline orientation deviation \(\sigma_{\theta}\) were varied in the simulations, as outlined in ~\ref{table1}. Each parameter was discretized into 10 steps, resulting in a total of 10,000 PF simulations of ferroelectric hysteresis.
\begin{table}[h!]
\centering
\resizebox{\columnwidth}{!}{
\begin{tabular}{|c|c|c|c|c|}
\hline
\textbf{Parameters} & \(D_{G}\) [nm] & \(\nu_{\text{ortho}}\) [a.u.] & \(\mu_{\theta}\) [rads] & \(\sigma_{\theta}\) [rads] \\ 
\hline
& 5 \(\sim\) 25 & 0.5 \(\sim\) 1 & 0 \(\sim\) \(\frac{\pi}{4}\) & 0 \(\sim\) \(\frac{\pi}{5}\) \\
\hline
\end{tabular}
}
\caption{Range of material parameters related to the polycrystalline structure used in the PF simulations, including average grain diameter \(D_G\), polar orthorhombic fraction \(\nu_{\text{ortho}}\), crystalline fiber texture \(\mu_{\theta}\), and crystalline orientation deviation \(\sigma_{\theta}\).}
\label{table1}
\end{table}

\subsection{Variational Autoencoder}

In this work, we employed a VAE to learn a compact, low-dimensional representation of ferroelectric hysteresis generated by PF simulations. A VAE is a generative model consisting of an encoder \(q_{\theta}(\boldsymbol{z}|\boldsymbol{x})\), which encodes the input sample \(\boldsymbol{x}\) into a latent representation \(\boldsymbol{z}\), and a decoder \(p_{\phi}(\boldsymbol{x}|\boldsymbol{z})\), which reconstructs the input from the latent space. Here, \(\theta\) and \(\phi\) are the parameters of the encoder and decoder, respectively. We specifically use a \(\beta\)-VAE \cite{betaVAE}, where the evidence lower bound (ELBO) is maximized during training:
\[
\mathcal{L}_{\theta,\phi}(\beta,\boldsymbol{x},\boldsymbol{z}) = \mathbb{E}_{q_{\phi}(\boldsymbol{z}|\boldsymbol{x})}[\log p_{\theta}(\boldsymbol{x}|\boldsymbol{z})] - \beta D_{\text{KL}}(q_{\phi}(\boldsymbol{z}|\boldsymbol{x}) || p(\boldsymbol{z}))
\]
The first term represents the reconstruction loss, computed using binary cross-entropy, while the second term is the Kullback-Leibler (KL) divergence between the approximate posterior \(q(\boldsymbol{z}|\boldsymbol{x})\) and the prior \(p(\boldsymbol{z})\), assumed to be a standard normal distribution \(\mathcal{N}(0,1)\). The hyperparameter \(\beta\) controls the trade-off between reconstruction quality and latent space regularization.

Both the encoder and decoder were implemented as multi-layer perceptrons (MLPs) with one hidden layer consisting of 256 units, and the latent dimension was set to 2. The \(\beta\) coefficient was set to 0.1, and the model was trained for 200 epochs with a batch size of 128 and a learning rate of \(10^{-3}\) using the Adam optimizer.

\subsection{Pearson Correlation Coefficient}

The Pearson correlation coefficient (\(r\)) is a statistical measure used to quantify the linear relationship between two variables. In this study, \(r\) is employed to evaluate the correlation between the VAE latent variables (\(Z_1, Z_2\)) and the material parameters or \(P\)-\(V\) loop characteristics. The Pearson correlation coefficient ranges from \(-1\) to \(1\), with values closer to \(-1\) or \(1\) indicating a stronger negative or positive linear relationship, respectively, and values near \(0\) suggesting no linear correlation. 
The coefficient \(r\) is calculated using the formula:  
\[
r = \frac{\text{Cov}(x, y)}{\sigma_x \sigma_y}
\]
where \(\text{Cov}(x, y)\) represents the covariance between the variables \(x\) and \(y\), and \(\sigma_x\) and \(\sigma_y\) denote the standard deviations of \(x\) and \(y\), respectively. 

\subsection{Gaussian Processes}

GPs are a nonparametric supervised learning method widely utilized for regression tasks. Here, the objective is to learn the function \(f\) that maps inputs \(\boldsymbol{x} = [Z_1, Z_2]\) (2D latent coordinates) to outputs \(\boldsymbol{y} = [D_G, \nu_{\text{ortho}}, \mu_\theta, \sigma_\theta]\) (material parameters), given a set of observations \(\boldsymbol{Y} = \{y_1, y_2, \dots, y_n\}\) at corresponding input points \(\boldsymbol{X} = \{\boldsymbol{x_1}, \boldsymbol{x_2}, \dots, \boldsymbol{x_n}\}\).

The GP prior over the latent function values \(f = \{f(\boldsymbol{x_1}), f(\boldsymbol{x_2}), \dots, f(\boldsymbol{x_n})\}\) is formally expressed as:

\[
f \sim \mathcal{N}(m(\boldsymbol{X}), K(\boldsymbol{X}, \boldsymbol{X}))
\]

where \(\mathcal{N}\) denotes a multivariate normal distribution, \(m(\boldsymbol{X})\) is the mean vector, and \(K(\boldsymbol{X}, \boldsymbol{X})\) is the covariance matrix with entries \(k(\boldsymbol{x_i}, \boldsymbol{x_j})\) defined by a kernel function. The radial basis function (RBF) kernel is chosen here, given by:

\[
k_{RBF}(\boldsymbol{x_i}, \boldsymbol{x_j}) = \sigma^2 \exp\left(-\frac{|\boldsymbol{x_i} - \boldsymbol{x_j}|^2}{2l^2}\right)
\]
where \(\sigma^2\) is the variance, and \(l\) is the length scale of the kernel. In addition to the RBF kernel, we used a White kernel, defined as:

\[
k_{W}(\boldsymbol{x_i}, \boldsymbol{x_j}) = \sigma_n^2 \delta_{ij}
\]

where \(\sigma_n^2\) represents the noise variance, which is set to 0.005, and \(\delta_{ij}\) is the Kronecker delta function. 
The kernel parameters were optimized during training using 1,000 randomly selected training points, comprising 2D latent encodings \(\boldsymbol{X} = [Z_1, Z_2]\) and their corresponding labels \(\boldsymbol{Y} = [D_G, \nu_{\text{ortho}}, \mu_\theta, \sigma_\theta]\). Notably, an independent GP was trained for the prediction of each label.

\section*{Data availability}

The data that support the findings of this study are available from the corresponding authors upon reasonable request.

\section*{Acknowledgements}

This work has received the financial support of IPCEI France 2030 Programs POI and eFerroNVM together with ANR-23-CE24-0015-01 ECHOES project.

\section*{Author Contributions}

Kévin Alhada-Lahbabi: Conceptualization (equal); Data curation (equal); Formal analysis (equal); Investigation (equal); Methodology (equal); Writing – review \& editing (equal).

Brice Gautier: Conceptualization (equal); Funding acquisition (equal); Resources (equal); Supervision (equal); Validation (equal); Writing – review \& editing (equal).

Damien Deleruyelle: Conceptualization (equal); Funding acquisition (equal); Resources (equal); Supervision (equal); Validation (equal); Writing – review \& editing (equal).

Grégoire Magagnin: Conceptualization (equal); Supervision (equal); Validation (equal); Data curation (equal); Formal analysis (equal); Investigation (equal); Methodology (equal); Project administration (equal); Writing – review \& editing (equal)

\section*{Competing Interests}
The authors declare no competing interests.

\nocite{langley00}

\bibliography{example_paper}
\bibliographystyle{unsrt}

\end{document}